# Tuning the graphene work function by electric field effect

Young-Jun Yu[1,2], Yue Zhao[1], Sunmin Ryu[3], Louis E. Brus[3],
Kwang S. Kim[2] and Philip Kim[1*]

[1]Department of Physics, Columbia University, New York, New York 10027, USA
[2] Center for Superfunctional Materials, Department of Chemistry, Pohang University of Science and Technology, Hyojadong, Namgu, Pohang 790-784, Korea
[3]Department of Chemistry, Columbia University, New York, New York 10027, USA
*E-mail: pk2015@ columbia.edu

**Abstract**

We report variation of the work function for single and bi-layer graphene devices measured by scanning Kelvin probe microscopy (SKPM). Using the electric field effect, the work function of graphene can be adjusted as the gate voltage tunes the Fermi level across the charge neutrality point. Upon biasing the device, the surface potential map obtained by SKPM provides a reliable way to measure the contact resistance of individual electrodes contacting graphene.

High conductivity[1,2] and low optical absorption[3,4] make graphene an attractive material for use as a flexible transparent conductive electrode[5-8]. This atomically thin carbon layer provides the additional benefit that its work function can be adjusted by the electric field effect (EFE). Since the band alignment of two different materials is determined by their respective work functions, control over the graphene work function is the key to reducing the contact barriers of graphene top electrode devices[9, 10]. Previous scanning probe based studies[11-13] reveal that the work function of graphene is in a similar range to that of graphite, ~4.6 eV[14], and depends sensitively on the number of layers[15, 16]. However, the active controlling of the graphene work function has yet to be demonstrated.

In this study, we apply Scanning Kelvin probe microscope (SKPM) techniques to back-gated graphene devices and demonstrate that the work function can be controlled over a wide range by EFE induced modulation of carrier concentration. SKPM is an atomic force microscope (AFM) based experimental technique that can map the surface potential variation of a sample surface relative to that of metallic tip[17]. The change of work function is ascribed by the Fermi level shift due to the EFE induced carrier doping and is well quantified by the electronic band structure of graphene. On biased graphene devices, SKPM also allows us to accurately measure graphene/metal contact resistances by mapping the surface potential of a device. The wide range of control over the work function demonstrated here suggests graphene as an ideal material for applications where work function optimization is important.

Graphene samples were prepared by mechanical exfoliation[18] on Si wafers covered with 300 nm thick $SiO_2$ and then Cr/Au electrodes (5 nm/30 nm thickness) were fabricated by standard electron beam lithography. The thickness of each graphene samples was characterized by Raman spectroscopy (see Supporting information, Fig. S1). In this work, we have studied with three single layer graphene (SLG) and two bilayer graphene (BLG) transistors. Fig. 1(a) shows a schematic diagram of the simultaneous SKPM experiments with the EFE induced carrier modulation. The SKPM measurements were performed by commercial atomic force microscope (XE-100, Park Systems Corp.) in air and dry nitrogen environment at room temperature. During observing SKPM data, we applied AC voltage amplitude of ~0.3-0.5 V and a frequency of 17 kHz to a Cr/Au coated probe. The SKPM images were obtained with two-way scan method to avoid topographic artifacts. The first scan was for topography by non-contact mode with dithering resonant frequency ~ 120-170 kHz and the second scan was for SKPM image with 10~30 nm constant height mode.

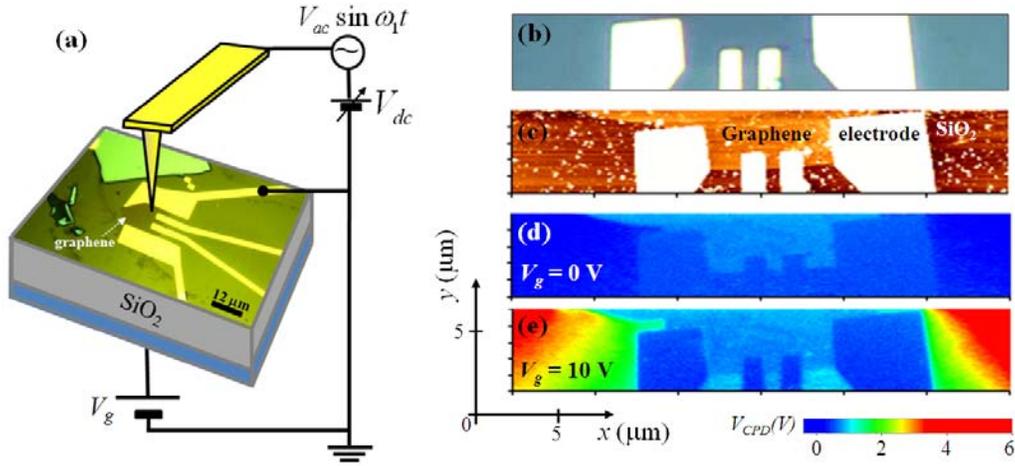

**Figure 1** (a) Schematic diagram for measuring the EFE modulation of the surface potential of graphene devices using the SKPM. Gate voltage $V_g$ is applied to the degenerately doped Si substrate and the electrodes of graphene device are grounded. (b)-(e) From top to down, optical image (b), AFM topography (c), and SKPM images of graphene device $V_g$ = 0 V (d) and 10 V (e), respectively.

The carrier density, and hence the Fermi energy, $E_F$, of the graphene, is controlled by the gate voltage $V_g$ applied to the degenerately doped Si substrate. Fig. 1(b)-(e) shows the optical, atomic force microscopy (AFM) topographic, and surface potential images of one of the SLG devices used in this experiment. In general, the local surface potential, $V_{CPD}$, obtained from the contact potential difference between the SKPM probe and local surface[17], is sensitively influenced by $V_g$. By comparing $V_{CPD}$ maps taken at $V_g$ =0 (Fig. 1(d)) and 10 V (Fig. 1(e)), we notice a much larger signal contrast at higher $V_g$. In particular, $V_{CPD}$ tends to increase to values comparable to $V_g$ on the insulating SiO$_2$ substrate. This indicates that the unscreened electric field from the back gate is the dominant source for the contrast in $V_{CPD}$ image in the bare SiO$_2$ regions. In order to minimize a direct exposure of the SKPM probe to this long range electrostatic influence, we limit our surface potential analysis to areas within the conducting part of the device, where the subtle local surface potential variation can be readily probed.

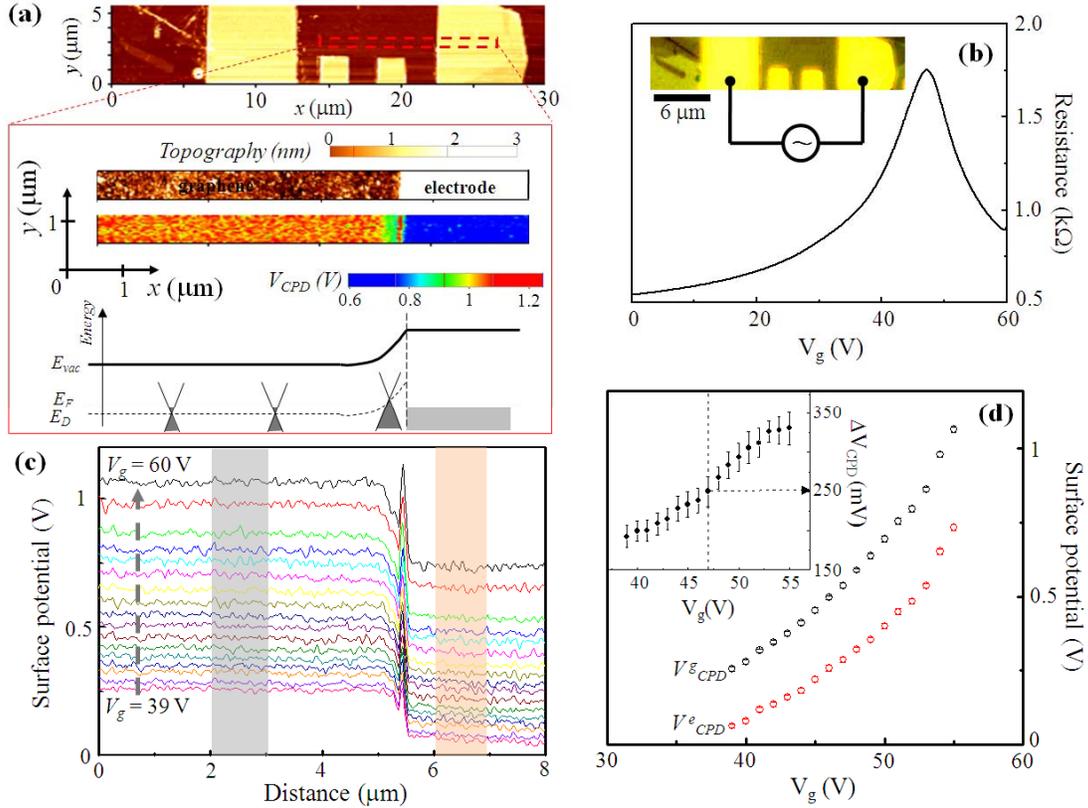

**Figure 2** (a) Top panel shows an AFM topography of the device. The surface potential map of a selected area far from the graphene edges (marked by a dotted line box) is analyzed in detail in order to minimze the effect from unscreened long ranged electrostatic force from the gate. Small spatial variations in the topographic and SKPM signals are found to be correlated on the top of the graphene surface. $V_g = 55$ V in this image. A schematic energy alignment diagram for the graphene sample and the metallic elecrode is displayed where $E_{vac}$, $E_F$, and $E_D$ are representing the vacuum energy level, Fermi energy and the charge neutrality point of graphene, respectively. (b) Transport measurement using the two outer terminals as shown in the inset. From the peak position we have $V_D = 48$ V. (c) The surface potential profile along the $x$-axis (averaged over the $y$-axis) at different gate voltages $V_g$ (at every 1 V from 39 to 60 V). (d) The CPD of graphene $V^g_{CPD}$ and electrode $V^e_{CPD}$ at different gate voltages obtained from the average surface potential in the selected gray and red shaded area in (c). The inset shows $\Delta V_{CPD} = V^g_{CPD} - V^e_{CPD}$ as a function of $V_g$. The vertical dashed line indicates $V_g = V_D$ where the corresponding $\Delta V_{CPD}$ is highlighted by the horizontal arrow.

Utilizing multi-terminal device geometry, we carry out a comparison study of the surface potential (Fig. 2(a)) and transport measurement (Fig. 2(b)) on the same device as a function of $V_g$. In the particular device shown in this figure, we determine the charge neutral gate voltage $V_D = 48$ V, from the gate voltage corresponding to a sharp resistance peak. In the surface potential map where $V_g$ is fixed, we observe a stepwise increase of $V_{CPD}$ at the junctions between graphene and electrodes regions, but no significant spatial variation within each region. Similar features are observed at different values of $V_g$ (Fig. 2(c)), where the overall line profiles of $V_{CPD}$ in the two regions shift upward with a common background as $V_g$ increases. This background signal is due to the unscreened long-range electrostatic interaction between the conducting cantilever probe and the back gate, and thus is insensitive to a small spatial position change of the SKPM tip. Since $V_g$ should not influence the local surface potential in metallic electrode, the back ground signal on the metal electrode surface can serve to separate the spatially constant background from the relative change in the local surface potential. In particular, the EFE induced local surface potential change in graphene, $\Delta V_{CPD}$, can be obtained by $\Delta V_{CPD} = V_{CPD}^g - V_{CPD}^e$, where $V_{CPD}^e$ and $V_{CPD}^g$ are the average $V_{CPD}$ in the electrode and the graphene, respectively (Fig. 2(d) inset). Interestingly, a sudden change $\Delta V_{CPD}$ is observed at the charge neutrality point ($V_g = V_D$), as indicated by the vertical dashed line.

Similar features are always present at the neutrality point in other SLG samples studied in this experiment. We also perform similar measurement of $\Delta V_{CPD}$ on BLG samples. Similar to the SLG, $\Delta V_{CPD}$ of BLG can be modulated by applied gate voltage (see Supplementary information, Fig. S2). Unlike the SLG samples, however, there is no such sudden change of $\Delta V_{CPD}$ at the BLG charge neutrality point, implying the difference of the electronic structures of the samples as we will discuss later.

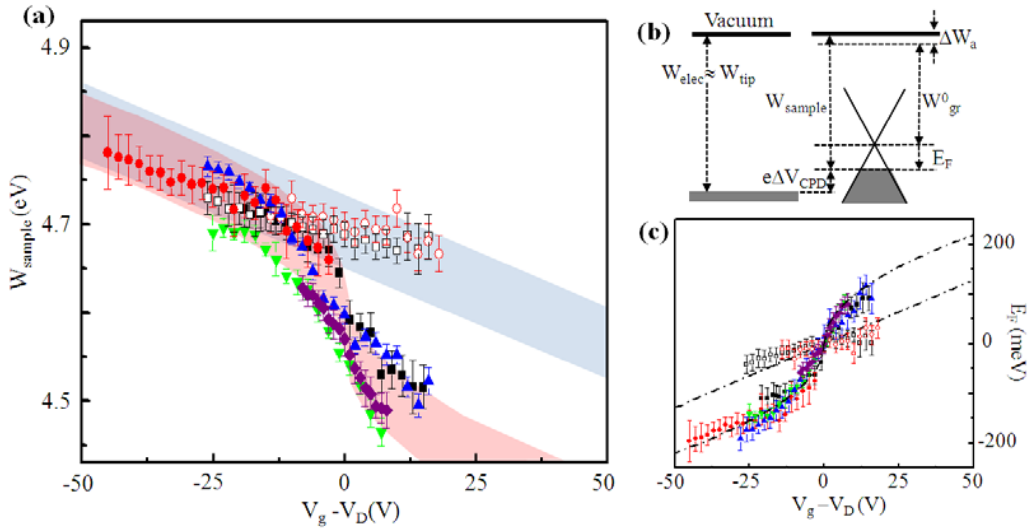

**Figure 3** (a) Measured work function of SLG samples (filled symbols) and BLG samples (open symbols) as a function of $V_g - V_D$. SLG samples show larger work function changes (shaded in red) while the BLG samples exhibit less changes (shaded in blue), where shaded areas indicate the uncertainty for the work function offset due to the adsorbate induced surface dipole layers (see supplementary materials). The filled green, red and purple symbols are SLG data taken in air. Other data were taken in dry nitrogen environment. (b) Schematic diagram for the energy level alignment in the SKPM tip and graphene samples. The left panel shows the work function of tip relative to the vacuum level. The right panel shows the relation between $W_{sample}$, $W_{gr}^0$, $E_F$, and $\Delta W_a$, defined in the text relative to the vacuum level. (c) The Fermi energy variation of SLG and BLG as a function of $V_g$-$V_D$. Symbols are same as in (a). $E_F$ is deduced from $W_{sample}$, by subtracting off $W_{gr}^0 + \Delta W_a$, whose values obtained at the charge neutrality point $V_g = V_D$. The dashed curve and line are the $E_F^{SLG}$ and $E_F^{BLG}$ from the band structure calculations.

The relative surface contact potential difference, $\Delta V_{CPD}$ obtained above can be related to the difference in work function between two different surfaces: $e\Delta V_{CPD} = W_{elec} - W_{sample}$, where $W_{elec}$ and $W_{sample}$ are work functions of the electrode and graphene sample surfaces, respectively. Since $W_{elec}$ is insensitive to $V_g$ due to the large density of states near the Fermi level in the gold electrodes, the observed variation of $\Delta V_{CPD}(V_g)$ reflects the EFE modulation of $W_{sample}$. We note that the work function of our gold coated tip is close to $W_{elec}$ since both the SKPM probe and the electrodes are made of gold and that they are exposed in the same experimental condition. A separately performed calibration measurement yields $W_{tip} = 4.82 \pm 0.08$ eV (see Supporting information, Fig. S3). Using this value, we estimate $W_{sample} \approx W_{tip} - e\Delta V_{CPD}$. Fig. 3(a) shows the resulting $W_{sample}$ as a function of $\Delta V_g = V_g - V_D$. Here we list results obtained from several SLG and BLG devices in different experimental conditions. For each data set, the transport characteristics are measured independently to obtain $V_D$. We found that $V_D$ ranges from 30 V to 50 V for most devices. The positive sign of $V_D$ indicates that the samples are hole doped from the environment, suggesting a dipole layer formation on the top of graphene surface. Despite this uncontrolled environment doping, we demonstrate that the work function of sample can be tuned by EFE within the range 4.5-4.8 eV for SLG and 4.65-4.75 eV for BLG in ambient and dry nitrogen conditions. With further optimization using chemcial functionalization, a wide range of work functions can potentially be achieved for applications which require adjustment of the work function.

The behaviour of EFE tuned $W_{sample}$ can be explained by the change of $E_F$ in graphene devices. From the schematic diagram in Fig. 3(b), we note $W_{sample} = \Delta W_a + W_{gr}^0 - E_F$, where $\Delta W_a$ is the offset of work function due to the adsorbate dipole layer formation, and $W_{gr}^0$ is the intrinsic work function of undoped graphene. In order to estimate $W_{gr}^0$ from the measured $W_{sample}$, an independent estimation of $\Delta W_a$ is necessary. In the experiments performed in air and dry nitrogen atmosphere, we estimated $|\Delta W_a| < 50$ meV for both SLG and BLG (see Supporting information, Fig. S4). This upper bound of $|\Delta W_a|$ allows us to estimate the work functions of undoped SLG and BLG from the measured $W_{sample}(\Delta V_g = 0)$. With this analysis, we obtained $W_{gr}^0 = 4.57 \pm 0.05$ eV for SLG and $4.69 \pm 0.05$ eV for BLG, which are in reasonable agreement with recent theoretical estimations[19-21]. Note that the work function of SLG is found to be smaller than that of BLG, indicating the chemical stability of BLG over SLG[22].

The controlled modulation of $W_{sample}$ allows us to estimate the EFE induced $E_F$ variation. Noting that $E_F(\Delta V_g = 0) = 0$ in our convention, we have $E_F(\Delta V_g) = W_{sample}(\Delta V_g = 0) - W_{sample}(\Delta V_g)$. Fig. 3(c) shows $E_F(\Delta V_g)$ calculated from the data sets displayed in Fig. 3(a). Remarkably, five data sets from SLG and two data sets from BLG all collapse into two separate families of the curves representing $E_F(\Delta V_g)$ for SLG and BLG, respectively. The Fermi energy variation of SLG and BLG can be described by the change of carrier density induced by the EFE, i.e., $E_F^{SLG} = sign(\Delta V_g)\hbar v_F \sqrt{\alpha \pi |\Delta V_g|}$ for SLG and $E_F^{BLG} = \hbar^2 \pi \alpha \Delta V_g / 2m^*$ for BLG, where $\alpha = 7.1 \times 10^{10}$ cm$^{-2}$V$^{-1}$ is the gate

capacitance in electron charge, $v_F =1\times10^6$ m/sec is the Fermi velocity of SLG, and $m^* = 0.033\ m_e$ is the effective mass of carrier in BLG relative to the bare electron mass $m_e$ obtained from the literature[23-25]. Employing these values, we plot $E_F^{SLG}$ (dashed curve) and $E_F^{BLG}$ (dashed line) in Fig. 3(c), where excellent agreements are found without any fitting parameters, indicating the observed work function variation is ascribed solely to the EFE modulation of $E_F$.

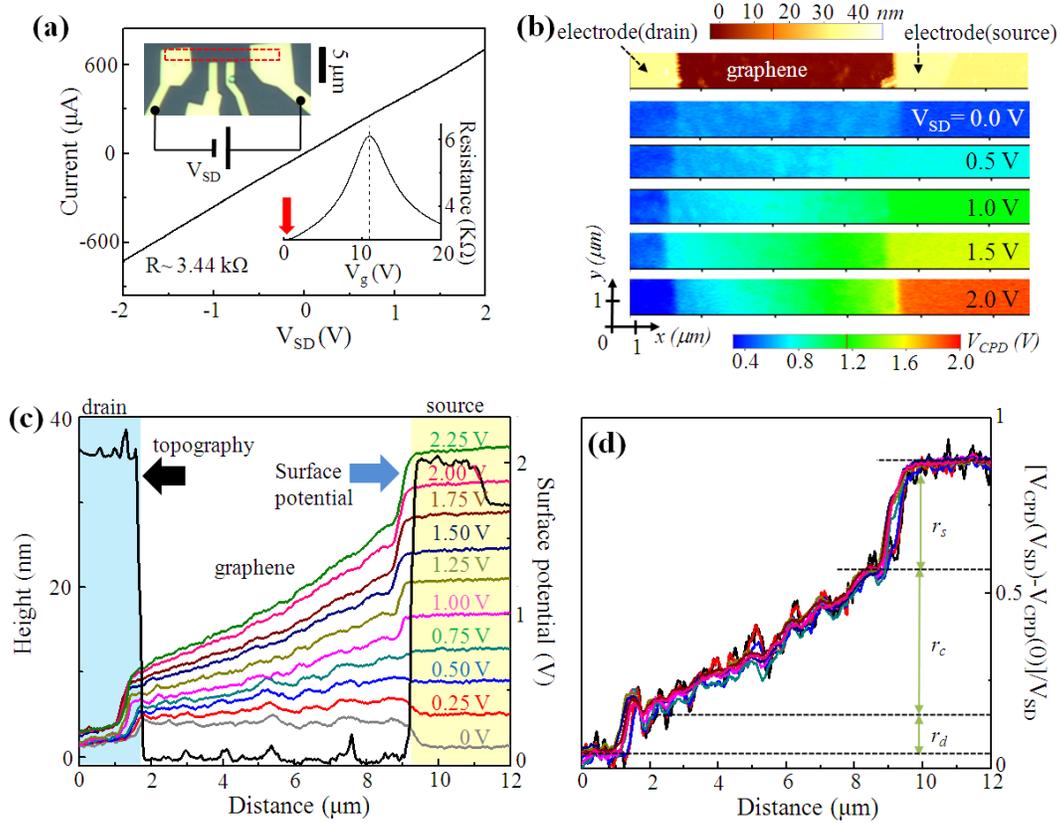

**Figure 4** (a) Current (*I*) and bias voltage ($V_{SD}$) characteristic of a SLG device. The slope of the curves indicates the two terminal resistance of device (3.44 kΩ). Upper inset shows the optical microscope image of the device. Lower inset shows the resistance as a function of $V_g$. The vertical arrow indicates the gate voltages and dotted line, $V_D$. (b) Topographic AFM image (top) and surface potential images of the area marked by the red rectangular box in (a). The drain electrode is grounded while the source electrode is biased by $V_{SD}$ as indicated in each panel. The gate voltage is fixed to $V_g = 0$. (c) Surface potential profiles along the *x*-axis shown in (b) at different $V_{SD}$. (d) Normalized and referenced surface potential profiles shown in (c). The surface potential referenced to the grounded drain electrode ($V_{CPD}(V_{SD}=0)$) and normalized by $V_{SD}$. All the curves shown in **c** are collapsed in a single curve, where distinct kinks appear due to the contact resistances between electrodes and graphene. The proportion of the source and drain contact resistances and the channel resistance are marked by $r_s$, $r_d$, and $r_c$, respectively.

Finally, we discuss the simultaneous SKPM surface potential mapping on graphene devices with a finite bias voltage $V_{SD}$ applied between two electrodes. Fig. 4(a) shows the current (*I*) versus $V_{SD}$ characteristics at a fixed gate voltage. The slope of *I*- $V_{SD}$ yields a resistance of 3.44 kΩ which includes the contributions of contact resistances between each electrode and the graphene channel. The simultaneous SKPM surface potential mapping provides a way to estimate these contributions independently. Fig. 4(b) displays topographic and surface potential mapping of the same device with increasing $V_{SD}$ from 0 to 2 V. The surface potential of the grounded source electrode remains close to 0 V, while the surface potential of the biased drain electrode shifts upward as $V_{SD}$ increases. From the horizontal profile of these images (Fig. 4(c)), we clearly observe linearly increasing $V_{CPD}$ in the channel and kinks in $V_{CPD}$ at the junctions. We attribute these sudden potential drops to the contact resistance between the electrodes and graphene. By taking the grounded drain electrode as the reference point ($V_{CPD}$ ($V_{SD}$=0)) and normalizing the $V_{CPD}$ by $V_{SD}$, all the surface potential profiles at different $V_{SD}$ collapse into one universal curve [$V_{CPD}$ ($V_{SD}$) - $V_{CPD}$ ($V_{SD}$ =0)] / $V_{SD}$ (Fig. 4(d)). The vertical ratios between the kinks and the slope correspond to the source and drain contact resistances ($r_s$ & $r_d$) and graphene channel resistance ($r_c$). Considering that the total two terminal resistance of this device is 3.44 kΩ, we obtain $r_d$:$r_c$:$r_s$=0.5: 1.7: 1.24 kΩ for this particular device.

In conclusion, by employing a gate modulated SKPM measurement, we have demonstrated that the work function of graphene can be substantially adjusted by EFE. This widely tunable work function makes graphene an attractive material for low contact barrier electrodes. The simultaneous SKPM surface potential imaging allows us to evaluate the contacts of graphene devices.

**Acknowledgements.** We thank Mark S. Hybertsen for helpful discussion. This work was supported by DOE (No. DEFG02-05ER46215, DE-FG02-98ER-14861), ONR (N000140610138), NSF NSEC (No. CHE-0117752), NYSTAR, FENA MARCO center, DARPA (FA8650-08-C-7838) through CERA program, KICOS(GRL), KOSEF (EPB Center: R11-2008-052-01000), and BK21(KRF).

**Supporting Information Available:** Optical images and Raman spectroscopy of single-layer graphene (SLG) and bi-layer graphene (BLG) for confirming thickness, the surface potential modulation of a BLG device by the change of gate voltage, the calibration of Work function of SKPM probes, Work function offset of SLG and BLG.

**Supplementary information**

**1. Confirmation the thickness of graphene sheets with Raman characterization**

In this experiment, we employ three SLG and two BLG devices. Raman spectroscopy (Fig. S1) is used to estimate the thickness of the graphene layers.

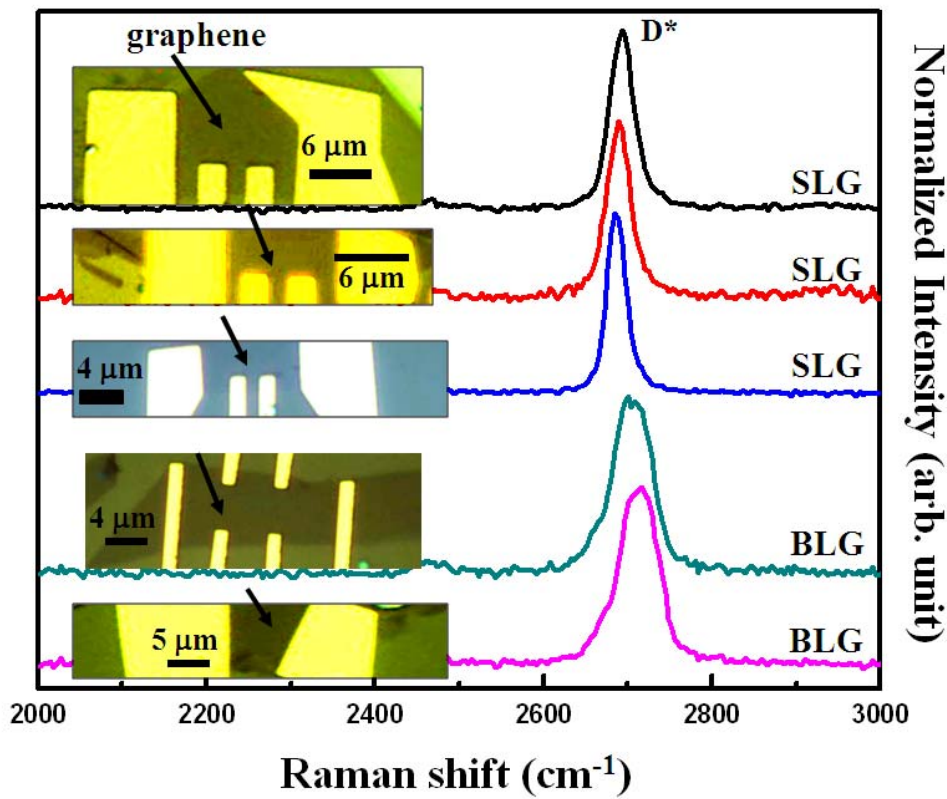

**Figure S1.** Insets show the optical images of three SLG and two BLG devices. The main panel shows $D^*$ resonance peaks which display the characteristic difference between SLG and BLG samples [S1]. 512 nm wave length laser is used for the excitation.

[S1] Ferrari, A. C., Meyer, J. C., Scardaci, V., Casiraghi, C., Lazzeri, M., Mauri, F., Piscanec, S., Jiang, D., Novoselov, K. S., Roth, S. & Geim, A. K. Raman spectrum of graphene and graphene layers. *Phys. Rev. Lett.* **97**, 187401 (2006)

## 2. Field Effect modulation of surface potential in bilayer graphene

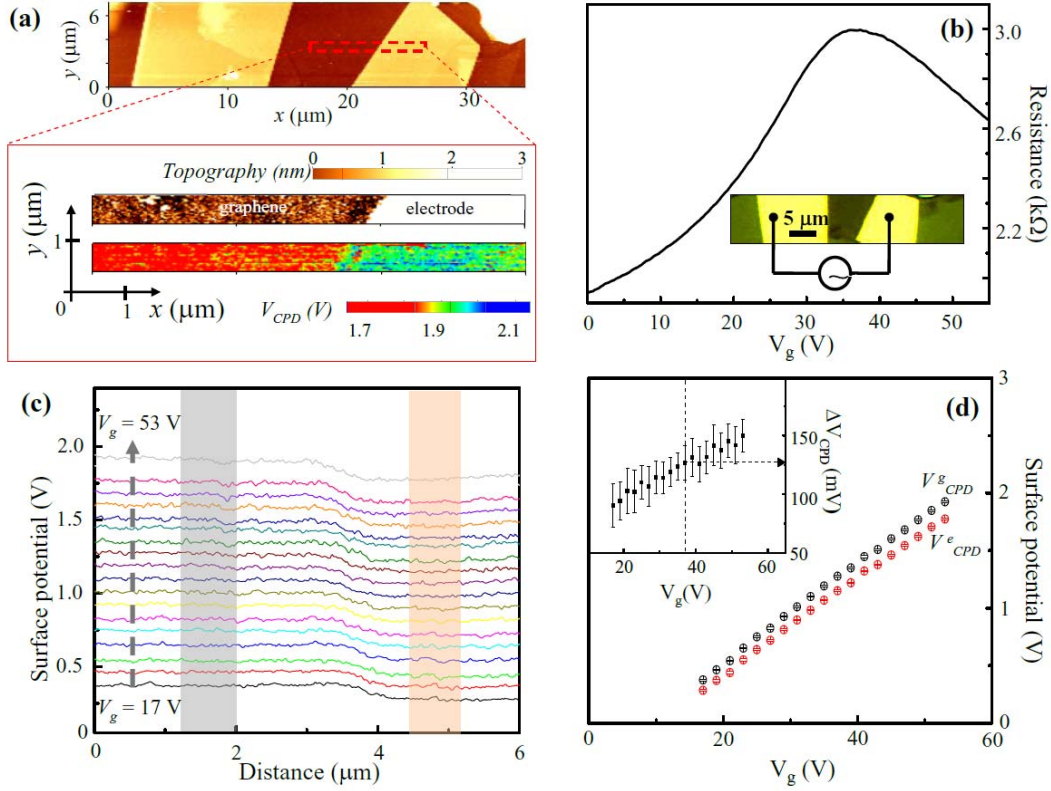

**Figure S2.** (a) Top panel shows an AFM topography of graphene channels with electrodes. The surface potential map of a selected area far from the graphene edges (marked by a dotted line box) is analyzed in detail in order to minimze the effect from unscreened long ranged electrostatic force from the gate. $V_g$ = 51 V in this image. (b) From the transport measurements using the two outer terminals as shown in the schematic diagram in the optical image inset, we obtain $V_D$ = 37 V. (c) Surface potential profile along the *x*-axis (averaged over the *y*-axis) at different gate voltages $V_g$ (at every 2 V from 17 to 53 V). (d) The CPD of graphene $V^g_{CPD}$ and electrode $V^e_{CPD}$ at different gate voltages obtained form the aveage surface potential in the selected gray and red filled area shown in (c). The inset shows $\Delta V_{CPD} = V^g_{CPD} - V^e_{CPD}$ as a function of $V_g$. The vertical dashed line indicates $V_g = V_D$ where the corresponding $V_{CPD}$ is highlighted by the horizontal arrow.

## 3. Calibration of SKPM probe with HOPG

In Fig.S3, the contact potential differences ($V_{CPD}$) between gold coated SKPM probes and highly oriented pyrolytic graphite (HOPG) are $V_{CPD}$ = -0.453, -0.132, -0.179 and -0.133 V. Thus, the averaged $V_{CPD}$ between SKPM probes and HOPG is -0.22 ± 0.08 V. The work function of the SKPM probes ($W_{tip}$) is estimated to be 4.82 ± 0.08 eV by subtracting $eV_{CPD}$ (=-0.22 ± 0.08 eV) from the work function of HOPG, $W_H$ (=4.6 eV) [S2, S3].

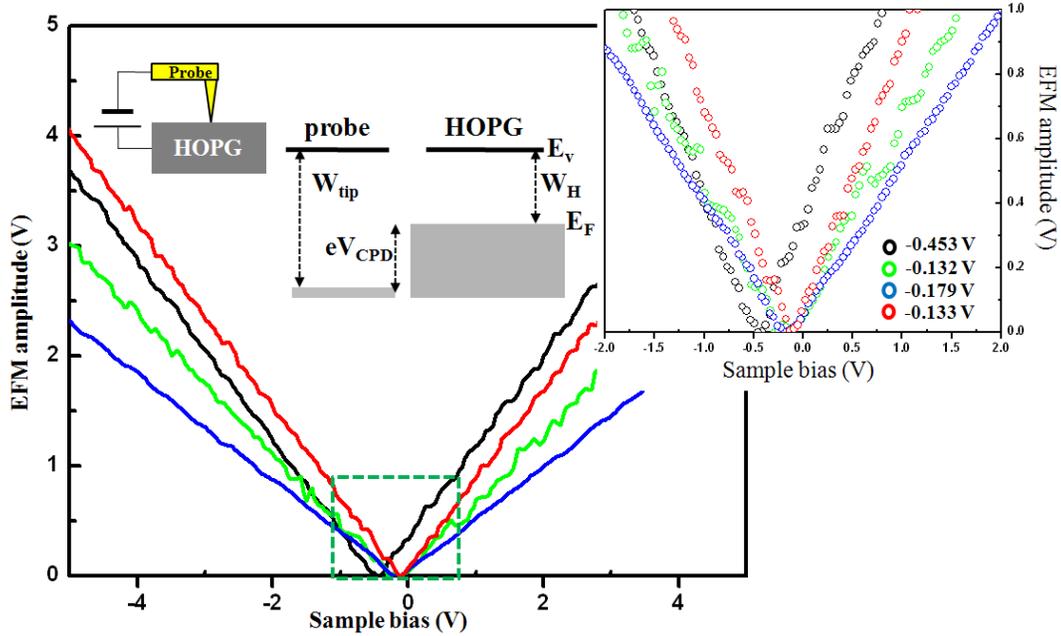

**Figure S3.** Variation of electric force microscope (EFM) cantilever amplitude is plotted as the function of the tip bias voltage relative to the HOPG substrate. The left figure shows a schematic diagram of experiment and the relationship between the Fermi energy and the work function of SKPM probe and HOPG. The right inset show the detailed plot of the zero bias regime in the green dotted box of the left figure. The sample is homogeneous and there is no back gate voltage to create a long range unscreened electrostatic force. Therefore we do not need to subtract off any electrostatic background signal in this tip calibration process

[S2] Takahashi, T., Tokailin, H. & Sagawa, T. Angle-resolved ultraviolet photoelectron spectroscopy of the unoccupied band structure of graphite. *Phys. Rev. B* **32**, 8317-8324 (1985)

[S3] Suzuki, S., Bower, C., Watanabe, Y. & Zhou, O. Work function and valence band states of pristine and Cs-intercalated single-walled carbon nanotube bundles *Appl. Phys. Lett.* **76**, 4007-4009 (2000)

## 4. Variation of the work function offset due to the adsorbate dipole layer formation

The range of the work function offset due to the adsorption of molecules from the environment can be estimated from the measured work function at the charge neutrality point, $W_{sample}(V_g=V_D) = W^0_{gr}+\Delta W_a$. Fig.S4 shows $W^0_{gr}+\Delta W_a$ as a function of $V_D$ for several SLG (filled symbols) and BLG (open symbols) samples. Despite different sample preparation and different environmental conditions of the experiment (i.e., air versus dry nitrogen), which results in a wide variation of $V_D$, the difference in work function is smaller than $\pm 50$ meV. Noting that $\Delta W_a$ depends sensitively on $V_D$, we thus estimate that $|\Delta W_a| < 50$ meV for the SLG samples. Since we do not see a specific trend of $\Delta W_a$ as a function of $V_D$, the work function of intrinsic graphene is obtained from the mean value of $W_{sample}(V_g=V_D)$, where we obtained 4.57 eV for SLG (black dotted line) and 4.69 eV for BLG (red dotted line).

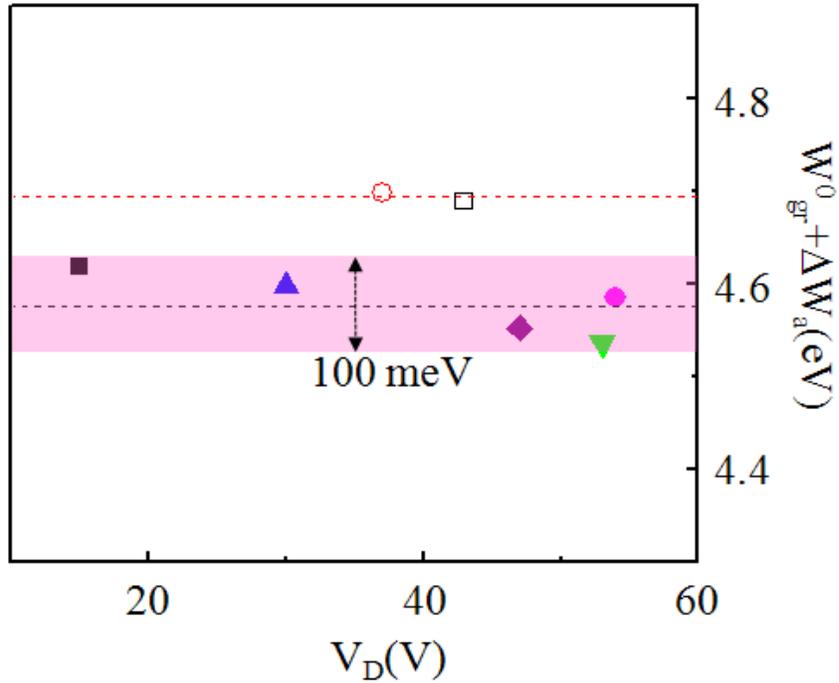

**Figure S4.** The work function ($W^0_{gr}+\Delta W_a$) of SLG (filled symbols) and BLG (open symbols) at the charge neutrality points ($V_D$) from Fig. 3(a). The shade in red denotes the deviation ($\pm 50$ meV) of work function offset ($\Delta W_a$) of SLG devices studied in this work. The black and red dotted line is the averaged work function ($W^0_{gr}$) of SLG (4.57 eV) and BLG (4.69 eV), respectively.